\documentclass[12pt,a4paper]{article}
\usepackage[utf8]{inputenc}

\usepackage{amsfonts,amssymb,amsmath,latexsym}
\usepackage{graphicx,graphics,epsfig}
\usepackage{hyperref}
\usepackage{xcolor}
\usepackage{float}
\usepackage{booktabs}
\usepackage{tabularx}
\usepackage{dcolumn}

\usepackage[T1]{fontenc}
\usepackage{tikz}
\usetikzlibrary{shapes.geometric, arrows}
\usetikzlibrary{arrows.meta, positioning}
\usepackage{comment}

\usepackage{chngcntr}
\usepackage{setspace}
\usepackage{multirow}%
\usepackage{amsthm}%
\usepackage[title]{appendix}%
\usepackage{textcomp}%
\usepackage{authblk}

\usepackage{adjustbox}
\usepackage{threeparttable}
\usepackage{tikz}
\usetikzlibrary{positioning}
\usetikzlibrary{arrows.meta}

\hypersetup{
    colorlinks=true,
    linkcolor=blue,
    filecolor=magenta,
    urlcolor=blue,
}
\begin{document}

\doublespacing

\title{The Emergency-Care Consequences of Disrupted Prevention: Evidence from Mammography Screening Pathway}

\author{
Moslem Rashidi\thanks{Corresponding author.\ Department of Economics, University of Bologna, Piazza Scaravilli, 40126 Bologna, Italy (email: moslem.rashidi2@unibo.it).} \quad
Luke Brian Connelly\thanks{Centre for the Business and Economics of Health, The University of Queensland, St Lucia, QLD 4072, Brisbane, Queensland, Australia (email: l.connelly@uq.edu.au), and Department of Sociology and Business Law, University of Bologna, Bologna, Italy (email: luke.connelly@unibo.it).} \quad
Gianluca Fiorentini\thanks{Department of Economics, University of Bologna, Piazza Scaravilli, 40126 Bologna, Italy (email: gianluca.fiorentini@unibo.it).}
}

\date{\today}

\maketitle

\begin{abstract}
Do disruptions to organized preventive-care pathways increase the likelihood of downstream overnight emergency hospitalizations? We study this question using the COVID-19 pandemic as a natural experiment. Using SHARE Wave~9 data on women aged 50--69 in eight European countries, we instrument for mammography uptake---an observable indicator of access to organized preventive-care pathways---with the interaction between country-level pandemic restriction intensity and SHARE interview-month cohort. This variation is plausibly exogenous because fieldwork timing shifted the portion of the 2020 restriction period that fell within each respondent's two-year recall window, generating differential disruption to screening access across cohorts within countries. The OLS estimates are close to zero, consistent with selection based on health status. However, the IV results imply that mammography uptake reduces the probability of overnight emergency hospitalization by approximately 6 percentage points among compliers. LIML produces a statistically significant estimate of $-0.114$. Women aged 70 and above, who are outside organized screening programs in all eight countries, show no first-stage and no reduced-form evidence of an effect on overnight emergency hospitalization. A decomposition exercise confirms that the breast-cancer detection channel accounts for at most 6 percent of the estimate, pointing instead to broader preventive-pathway disruption.
\end{abstract}

\begin{center}

\bigskip


\end{center}

\vfill

\thispagestyle{empty}

\section{Introduction}
\label{sec:introduction}
Do disruptions to organized preventive and primary-care services increase subsequent overnight emergency hospitalization? The question matters for how we evaluate the economic returns to preventive care and for how we assess the systemic costs of health-care crises. Organized screening programs are not stand-alone diagnostics. They are institutional entry points to a broader clinical network: women who enter an organized mammography program are identified, referred for follow-up imaging when needed, placed into specialist pathways, and kept under planned management. Disrupting that entry point, therefore, does more than delay a single test. It can break a chain of care that would otherwise have kept certain conditions within planned, rather than emergency, settings. The economic case for preventive care rests partly on precisely this substitution: earlier detection and coordinated follow-up may reduce downstream emergency or late-stage hospital use. If the substitution is quantitatively meaningful, organized programs may generate fiscal returns well beyond their direct health benefits. If it is weak, the case is correspondingly thinner. This paper estimates the substitution using the COVID-19 disruption to organized mammography programs across eight European countries.

We study this question with mammography uptake serving as an observable marker of the broader disruption to organized preventive and primary-care services during the pandemic. Specifically, in our conception, unobserved disruptions to preventive care more generally are captured by the COVID-19 restrictions, which simultaneously disrupted a wide range of preventive and primary-care activities: mammography programs were suspended, GP surgeries reduced non-urgent appointments, referral pathways were interrupted, and patients avoided clinical settings because of concerns about infection. These disruptions cannot, however, all be directly observed in SHARE. What we can observe is mammography uptake, a well-measured preventive-care variable embedded in exactly the organized referral and follow-up infrastructure we hypothesize was disrupted. Mammography uptake, therefore, acts as a barometer of broader preventive-care disruption. The IV estimate we recover is not the isolated biological effect of a missed mammogram. It is the downstream acute-care consequence of disruption to organized preventive pathways, scaled by the measurable component of that disruption.

The empirical challenge is familiar. Women who comply with mammography recommendations differ from non-compliers in ways that also predict subsequent hospital use. The direction of this selection is theoretically ambiguous. Healthier, more prevention-oriented women may be both more likely to screen and less likely to be hospitalized, biasing OLS estimates downward. Women with greater perceived health risk may be more likely both to screen and to present to emergency care when symptoms appear, biasing OLS upward. As Einav et al. (2020) emphasize, individuals who comply with preventive-care recommendations can have health profiles that differ systematically from both self-selectors and never-takers. Ordinary least squares comparisons of screened and unscreened women are therefore unlikely to recover the causal effect of interest. In our data, the OLS estimate is close to zero, consistent with these competing selection forces roughly cancelling out and obscuring whatever causal relationship exists.

We use microdata from the Survey of Health, Ageing and Retirement in Europe (SHARE) Wave~9, covering women aged 50--69 in eight countries with population-based mammography programs. The identification strategy exploits the fieldwork structure of Wave~9, which ran from March through August 2022. Respondents report whether they had a mammogram during the previous two years. As a result, each respondent's recall window begins in a different calendar month in 2020, exactly when first-wave COVID-19 restrictions disrupted organized screening across Europe. A woman interviewed in March 2022 has a recall window beginning in March 2020, when restrictions were at their peak. A woman interviewed in August 2022 has a recall window beginning in August 2020, when restrictions had eased in most countries. We assign each woman the restriction index prevailing in her country at the month her two-year recall window opened. Conditional on interview-month fixed effects, the identifying variation comes from cross-country differences in restriction intensity at the start of the recall window among women interviewed in the same month.
 
This design does not compare women who chose to screen with women who chose not to screen. Instead, it compares women whose opportunity to comply with organized screening was differentially disrupted by country-level restrictions at the beginning of the recall period. The relevant exclusion restriction is that, conditional on interview timing and observed controls, early-pandemic restriction intensity affected later emergency hospitalization through its effect on mammography uptake rather than through other persistent consequences of the pandemic. This is the central identifying assumption. We assess it using controls for contemporaneous health-system strain and individual pandemic conditions, placebo populations outside the organized screening age range, and a pre-pandemic falsification exercise.
 
The preferred two-stage least squares estimate is $-0.060$ with a standard error of $0.028$. Interpreted as a local average treatment effect, this estimate implies that mammography uptake reduced the probability of an overnight emergency hospitalization by about 6 percentage points for women whose screening behavior was shifted by the disruption. The magnitude is large relative to the sample mean of 3.7 percent, so we interpret it as a local effect for the screening-margin population rather than as the average effect of mammography for all eligible women. The contrast with the OLS estimate of $-0.005$ is consistent with substantial selection into screening. Because the instrument contains multiple excluded components, we also report limited-information maximum likelihood estimates as a robustness check. LIML gives a larger estimate, $-0.114$ with a standard error of $0.056$, and the confidence interval overlaps the 2SLS result.

Several pieces of evidence support the interpretation of the IV estimates. First, the specifications control for health-system strain, economic stress, mental distress, and COVID-related burden measured in the first SHARE Corona Survey, which reduces the concern that restriction intensity only proxies for broader pandemic conditions. Second, the timing of the design separates the instrument period from the outcome period: restrictions are measured between March and August 2020, while emergency hospitalization is measured during the twelve months before the Wave~9 interview, so at least seven months separate the end of the instrument period from the beginning of the outcome window. This reduces the concern that the estimates are driven mechanically by contemporaneous hospital capacity constraints. Third, women above age 70 fall outside the organized screening target range in all eight countries. Following the negative-control logic of Danieli et al. (2026), we show that the instrument does not predict emergency hospitalization or other placebo outcomes in this group. Finally, replicating the full IV strategy in pre-pandemic SHARE Wave~8 produces estimates that are near zero or positive rather than negative. The pattern is consistent with an organized-pathway mechanism: the effect appears among women for whom the screening pathway operates and is absent where that pathway should not be affected.

The paper contributes to three literatures. The first is work on the returns to preventive care. Chandra, Gruber, and McKnight (2010) and Currie and Gruber (1996) study preventive-to-acute substitution using demand-side variation in cost-sharing and insurance coverage. We study the same economic margin using a supply-side disruption to an organized screening pathway. The second is the literature on healthcare access and acute-care use. Card, Dobkin, and Maestas (2009) and Finkelstein et al. (2012) show that access to care can change downstream emergency utilization. Our setting isolates a specific preventive-care pathway rather than access to health care in general. The third is the literature on COVID-19 healthcare disruptions. Ng and Hamilton (2022) and Li et al. (2023) document sharp declines in mammography volumes during 2020, while Maringe et al. (2020) project downstream mortality from diagnostic delays using assumed stage-progression rates. We complement this evidence by estimating realized downstream emergency hospitalization outcomes after the disruption.

The paper proceeds as follows. Section~\ref{sec:Methods} describes the data, sample, and identification strategy. Section~\ref{sec:Results} presents the main estimates and robustness checks. Section~\ref{sec:Discussion} discusses the mechanism and implications.

\section{Data and Methods}
\label{sec:Methods}

We use version 9.0.0 of the Survey of Health, Ageing and Retirement in Europe (SHARE), a cross-country panel survey of older Europeans. Our main analysis draws on Wave~9 (2021--2022), which covers 46,161 households and 69,447 individuals in 27 European countries and Israel. We also use the first SHARE Corona Survey, a special telephone survey fielded between June and August 2020 among SHARE panel participants aged 50 and above (Bergmann and Borsch-Supan, 2021; Bergmann et al., 2024), to capture respondents' circumstances during the initial phase of the pandemic.

We restrict the sample to respondents living in residential households, excluding those in nursing homes or other institutions. SHARE sampling frames differ across countries and coverage of institutionalized populations is limited (Bergmann et al., 2017). We also exclude respondents who appear in SHARE only as spouses or partners of age-eligible participants, as they are not representative of the target population.

We identify women eligible for organized mammography screening using the European Commission report \textit{Cancer Screening in the European Union}. The sample is restricted to women aged 50--69, the age group targeted by population-based breast cancer screening programs in most participating countries. We exclude countries without a fully implemented program (Bulgaria, Greece, Romania, and Slovakia), countries not covered by the report (Israel and Switzerland), countries with screening intervals longer than two years (Malta), and respondents who reported a breast cancer diagnosis in Wave~8 and were reinterviewed in Wave~9. The final sample covers eight countries: Austria, Belgium, the Czech Republic, Denmark, Germany, Poland, Slovenia, and Spain.

Our primary outcome is a binary indicator for having only emergency hospital stays in the last 12 months, constructed from the SHARE Wave~9 health-care module. We first classify respondents by whether they had any overnight hospital stay in the previous year and, among those who did, whether the stays were emergency or planned. The outcome equals 1 for respondents whose overnight hospital use consists exclusively of emergency stays and 0 otherwise -- that is, no overnight stay or at least one planned stay (Figure~\ref{fig:flow_RERonly}).

We use all-cause rather than cause-specific emergency hospitalization as our primary outcome. SHARE does not record admitting diagnoses for emergency stays, which precludes a cause-specific measure. Beyond this constraint, all-cause hospitalization is the substantively appropriate measure: organized screening disruption can generate overnight emergency hospitalization through several channels, including delayed detection of screen-identified pathology and foregone GP contact that the screening referral pathway would otherwise have triggered. An all-cause measure captures the total health system burden of preventive care disruption. Table~\ref{tab:SUMMARY_STAT_Y} reports summary statistics for this variable; the sample mean is 0.037, indicating that approximately 3.7\% of respondents experienced an emergency overnight hospitalization in the past year.

\begin{table}[ht]
\caption{Summary statistics of emergency overnight hospitalization in the past year}
\begin{center}
\begin{tabular}{lrrrrrr}
\hline
\multicolumn{1}{l}{}&
\multicolumn{1}{c}{}&
\multicolumn{1}{c}{}&
\multicolumn{1}{c}{}&
\multicolumn{1}{c}{}&
\multicolumn{1}{c}{}\\
\multicolumn{1}{l}{Variable}&
\multicolumn{1}{c}{Obs.}&
\multicolumn{1}{c}{Mean}&
\multicolumn{1}{c}{SD}&
\multicolumn{1}{c}{Skewness}&
\multicolumn{1}{c}{Kurtosis}\\
\hline
Emergency overnight hospitalization      &          2332 &     .037&  .1895 &    4.88&    24.84 \\
\hline
\end{tabular}
\parbox{120mm}{\footnotesize {\em }
}
\end{center}
\label{tab:SUMMARY_STAT_Y}
\end{table}

\subsection{Choice of Predictors}
\label{sec:predictors}

Our main explanatory variable is mammogram uptake, measured in SHARE Wave~9. Respondents were asked whether they had undergone a mammogram within the past two years. We define a binary indicator equal to 1 if the respondent reports a mammogram within that period and 0 otherwise.

All specifications control for a standard set of demographic and socioeconomic characteristics: age and age squared, educational attainment, living with a partner, household size, supplementary health insurance, and health literacy. Education follows the ISCED 1997 classification, collapsed to a binary indicator equal to 1 for levels 4--6 (upper secondary and above) and 0 for levels 0--3. Health literacy is based on whether respondents report needing help reading written information from doctors or pharmacies; we define high health literacy as never needing such help.

We also include six variables from the first SHARE Corona Survey to control for pandemic-related shocks that may jointly affect mammography uptake and hospital use. Two integer indices measure healthcare strain: one covering hospital-based problems (long waits, crowding, staff pressure, supply shortages, inadequate infection-prevention measures) and one covering non-hospital care. A third variable indicates whether the respondent reports any forgone, postponed, or denied medical care since the outbreak. Economic stress is captured by an index of financial support receipt, difficulty making ends meet, postponed bill payments, and use of savings for basic expenses. Mental distress is a standardized index of anxiety, depression, sleep problems, and loneliness. Finally, a COVID-burden indicator equals 1 if at least one close contact experienced COVID-like symptoms, tested positive, was hospitalized, or died from COVID-19.

All specifications include country fixed effects to absorb time-invariant cross-country differences in health systems, screening institutions, and public policies. Descriptive statistics for all covariates are reported in Table~\ref{tab:descriptives_main} in Appendix~\ref{Appendix B: Table}.

\subsection{Instrumental Variable Strategy}
\label{subsec:IV}

Identifying the causal effect of mammography on emergency hospitalization 
requires a source of variation in screening uptake that is unrelated to the 
individual health characteristics, preferences, and access conditions that 
drive both screening decisions and hospitalization risk. We exploit a measurement feature of SHARE Wave 9 that generates exactly this variation.

In Wave 9, respondents were asked whether they had a mammogram in the 
previous two years, with the recall window running backward from the month 
of interview. Interviews were conducted between March and August 2022. The 
start of each respondent's two-year recall window, therefore, falls in 
March--August 2020, the period when COVID-19 restrictions most severely 
curtailed non-urgent care across Europe. Crucially, because interview timing 
differed across respondents, so did the calendar month that opened their 
recall window. A woman interviewed in March 2022 has a recall window 
beginning in March 2020, when restrictions were at their most severe. A woman 
interviewed in August 2022 has a recall window beginning in August 2020, 
when restrictions had eased considerably in most countries. Women interviewed 
earlier were therefore more heavily exposed to the peak disruption period at 
the start of their reporting window, which mechanically reduced their 
probability of having a mammogram within the recall window.

Any systematic differences across interview cohorts that are common to all countries --- including the possibility that health-conscious respondents complete the survey earlier or later within a wave --- are absorbed by interview-month fixed effects, which we include in all specifications. Conditional on these fixed effects, the source of identifying variation is cross-country differences in restriction intensity at the recall-window start month, within each interview cohort. This is a country-level policy variable that no individual woman chose or could have anticipated at the time of her screening decision. The instrument assigns each respondent the restriction index prevailing in her country at the moment her recall window began:
\begin{equation}
Z_{icm} = R_{c,\,t(m)},
\label{eq:instrument}
\end{equation}
where $R_{c,t}$ measures restriction intensity in country $c$ during month 
$t \in \{\text{March},\ldots,\text{August}\;2020\}$, and $t(m)$ maps each 
interview month $m$ to the corresponding recall-window start month in 2020. 
Women interviewed in different months and living in countries with different 
restriction regimes were differentially exposed to pandemic-related disruption at the start of their recall window. This cross-country, cross-cohort variation is the source of identification.

We restrict the sample to women interviewed between March and August 2022, 
so that every respondent's recall window begins within the first wave 
disruption period. This yields six interview cohorts, each corresponding to 
a distinct recall-window start month in 2020. The two-stage least squares 
system is:
\begin{align}
D_{icm} &= \pi Z_{icm} + X_{icm}'\gamma + \alpha_c + \lambda_m + \nu_{icm},
\label{eq:first_stage}\\[4pt]
Y_{icm} &= \beta\,\widehat{D}_{icm} + X_{icm}'\delta + \alpha_c + \lambda_m 
+ \varepsilon_{icm},
\label{eq:second_stage}
\end{align}
where $D_{icm}$ is an indicator for mammography uptake in the previous two 
years, $Y_{icm}$ is an indicator for overnight emergency hospitalization in 
the 12 months preceding the interview, $X_{icm}$ is a vector of individual 
and household controls, $\alpha_c$ denotes country fixed effects, and 
$\lambda_m$ denotes interview-month fixed effects. Equation~\eqref{eq:first_stage} 
is the first stage: it estimates the restriction intensity at the start of the recall window, shifts mammography uptake, conditional on controls and 
fixed effects. Equation~\eqref{eq:second_stage} is the second stage: it 
estimates the effect of the instrumented mammography uptake on emergency 
hospitalization.

With both sets of fixed effects included, identification comes from 
cross-country differences in restriction severity during March--August 2020, 
aligned with the start of each respondent's recall window. Later portions of 
the recall window --- September 2020 through early 2022 --- are common across 
all six cohorts. By early 2022, public-health restrictions across Europe had 
been substantially relaxed relative to the first wave (WHO Regional Office for Europe, 2022; European Centre for Disease Prevention and Control, 2022), so later pandemic developments generate little differential variation in mammography uptake across cohorts once controls and fixed effects absorb the shared trends. The identifying variation is therefore concentrated in the March--August 2020 period, where cohorts diverge in their restriction exposure.

\begin{figure}[htbp]
	\centering
		\includegraphics[width=4.9in]{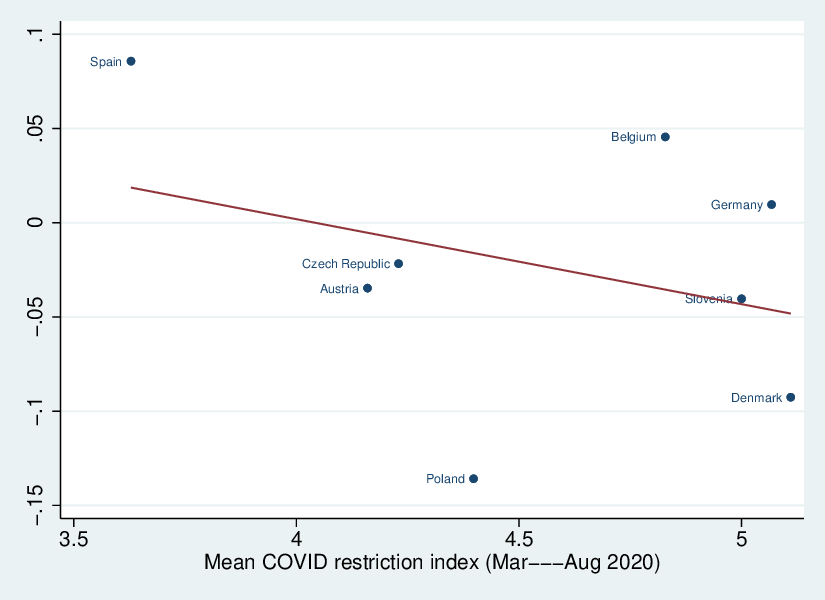}
\caption{Country-level mean restriction instrument $Z_c$ and mammography screening rates, Wave 9 vs.\ Wave 8}
	\label{Fig:RM_CC}
\end{figure}

The first requirement for a valid instrument is relevance: stricter 
restrictions must have reduced mammography uptake for at least some 
respondents. This is documented directly in the first-stage results in 
Section~\ref{sec:Results}. The cluster-adjusted first-stage $F$-statistic 
is 297.5, far above conventional thresholds for weak-instrument concern, and 
the partial $R^2$ is 0.104. Figure~\ref{Fig:RM_CC} illustrates the 
relationship at the country level: countries whose respondents faced higher average restriction intensity during March--August 2020 show lower Wave 9 
mammography rates and larger declines relative to Wave 8. Relevance is not 
in question.

For a local average treatment effect interpretation, monotonicity requires that greater restriction exposure did not increase mammography uptake for any woman while decreasing it for others. This is satisfied by the construction in our setting. COVID-19 restrictions reduced screening opportunities through program suspension, facility closures, and fear of infection. They did not create new pathways to mammography. The assumption, therefore, rules out only the implausible case in which some women sought screening specifically because restrictions were in place. Under monotonicity, the 2SLS estimator identifies the average treatment effect for compliers: women who attended organized screening under normal conditions but did not because pandemic restrictions curtailed the program. Appendix~\ref{sec:LATE} provides a formal treatment of the complier structure under a multi-valued instrument.

The exclusion restriction requires that $Z_{icm}$ affects emergency hospitalization $Y_{icm}$ only through its effect on access to organized preventive-care pathways, proxied by mammography uptake $D_{icm}$, conditional on controls and fixed effects. Under the reframing we adopt, this requires some care in interpretation. COVID-19 restrictions disrupted a broad range of preventive and primary-care services, not mammography alone. The exclusion restriction should therefore not be read as claiming that restrictions affected emergency hospitalization exclusively through the single clinical act of obtaining a mammogram. The argument is rather that restrictions affected emergency hospitalization through the organized preventive-care pathway, of which mammography is the observable first-stage indicator. That pathway includes the mammogram itself, the follow-up imaging and specialist referral it can trigger, the GP contact associated with screening visits, and the broader clinical infrastructure that a functioning, organized screening program activates. Mammography uptake is the measurable component of disruption to this full pathway. Women who lost access to the organized screening program lost access to the entire clinical chain, not merely to a single imaging procedure. The IV estimate therefore scales the downstream acute-care consequence of the broader disruption by its observable first-stage indicator; it is not the pure biological or clinical effect of a mammogram. This interpretation is supported, rather than threatened, by the decomposition exercise in Table~\ref{tab:bc_bounds}: because the breast-cancer emergency channel alone accounts for at most a small fraction of the estimated magnitude, the result is most naturally read as reflecting disruption to the organized preventive-care pathway as a whole. The exclusion restriction is plausible under this framing, but it cannot be verified directly. Four arguments and tests support its plausibility.

\textit{Temporal separation.} The instrument is measured in March--August 
2020. For a respondent interviewed in March 2022, the hospitalization window began in March 2021, at least seven months after the instrument period ended. This separation rules out any contemporaneous effect of restrictions on hospitalization and limits the scope for direct paths from the instrument to the outcome.

\textit{Nature of the outcome}. Overnight emergency hospitalizations are by definition unplanned. Unlike elective care, they cannot be scheduled or 
deferred in response to the restricted environment underlying the instrument. Direct effects of the instrument on the outcome through planned utilization decisions are therefore implausible.

\textit{Controls for observed alternative channels.} The most credible alternative path from restriction intensity to emergency hospitalization runs through broader pandemic disruption: healthcare system strain, economic stress, and mental distress, all of which were elevated during the first wave and could independently affect later hospital use. We control directly for each of these using the first SHARE Corona Survey module, fielded between June and August 2020. The IV estimates reported in Section~\ref{sec:Results} are therefore identified net of these observed channels. An alternative path through restriction intensity would have to operate through unobserved pandemic shocks that are orthogonal to the observed controls, a residual threat that is less plausible.

\textit{Falsification diagnostics.} We implement two pre-specified 
falsification tests. Following Danieli et al. (2026), we use women aged 70 
and above as a placebo group. Women over 70 fall outside the organized 
screening target range in all eight countries, so the instrument has no 
first-stage relevance through the mammography pathway in this group. Any 
association between the instrument and outcomes in this population, therefore, reflects an alternative path $Z \rightarrow U \rightarrow Y$ rather than the treatment channel. We test whether three pre-specified negative-control outcomes --- emergency hospitalization, dental visits, and influenza vaccination --- are predicted by the instrument in this placebo group, conditional on controls and fixed effects. They are not: the joint $F$-test fails to reject the null, and the incremental $R^2$ from adding the negative controls is negligible. We also replicate the design on pre-pandemic Wave 8 data, where COVID-19 could not have disrupted screening. No negative effect emerges. The Wave 9 result, therefore, reflects a pandemic-specific mechanism rather than a spurious feature of interview timing or seasonal patterns in healthcare use. Full details of both exercises are in Appendix~\ref{subsec:negcontrols_apv} and Appendix~\ref{subsec:falsification_iv_w8}. 

\subsection{Empirical Method}
\label{sec:Empirical method}

We estimate the effect of mammography uptake on overnight emergency
hospitalization using 2SLS as the primary estimator, with LIML as a
robustness check motivated by finite-sample concerns specific to this
setting.

Under the identification assumptions in Section~\ref{subsec:IV}, 2SLS
consistently estimates the LATE of mammography on emergency
hospitalization for complier subgroups. In this setting, however, the instrument structure raises a specific finite-sample concern:
many-instrument bias.

Our instrument assigns each respondent a restriction index based on
her country and interview cohort, producing six cohort indicators
interacted with eight country indicators as excluded instrument
components. In settings with many instruments, 2SLS is biased toward
the OLS probability limit, and the bias worsens as the instrument
count grows relative to sample size regardless of first-stage strength
(Bekker, 1994; Hansen, 2022). LIML corrects this bias by jointly
maximizing the likelihood of both stages. It is median-unbiased in a
wider class of many-instrument settings and consistent under the same
assumptions as 2SLS. If LIML and 2SLS produce similar estimates, the
many-instrument bias concern is minor; if they diverge, finite-sample distortion in the 2SLS result is the likely cause.

Standard errors are clustered at the country level. The instrument
assigns restriction intensity at the country-month level, so all women
in the same country interviewed in the same month share the same
instrument value. Country-level clustering allows for arbitrary
residual correlation within countries across all individuals and time
periods. With eight clusters, the asymptotic justification for
cluster-robust inference is strained and the resulting $p$-values
should be read with caution. We therefore weight the consistency of
sign and magnitude across both IV estimators more heavily than the
significance level from any individual specification.

\section{Results}
\label{sec:Results}

Table~\ref{tab:mammography_iv} reports OLS and instrumental variable
estimates of the effect of mammography uptake on overnight emergency
hospitalization in the last 12 months.

The OLS coefficient on mammography is $-0.005$ (standard error 0.007), small in magnitude and statistically indistinguishable from zero. This near-zero correlation is not surprising. Women who attend organized screening differ systematically from those who do not: they tend to be healthier, more health-literate, and better connected to primary care. These characteristics lower their baseline risk of emergency hospitalization, generating a downward bias in OLS that works against finding a negative effect. At the same time, women with higher unobserved health risk may be more motivated to attend screening, generating an upward bias in the opposite direction. The net result of these competing selection forces is a coefficient near zero that is uninformative about the causal effect of mammography.

Table~\ref{tab:first_stage_summary} reports summary statistics for the first-stage regression of Wave~9 mammography uptake on the excluded instrument components and the full control set. The cluster-adjusted first-stage $F$-statistic is 297.497 ($p < 0.001$) and the partial $R^{2}$ is 0.104. Both indicate strong instrument relevance: the excluded components explain a substantial share of residual variation in screening participation after controls and country fixed effects are absorbed. The robust $R^{2}$ of 0.024 reflects variation explained after accounting for within-country clustering. By conventional thresholds, weak identification is not a concern in this setting.

The 2SLS coefficient on mammography uptake is $-0.060$ (standard error $0.028$, $p < 0.05$). Under the identification assumptions, this estimate captures the downstream acute-care consequence of COVID-induced preventive-care disruption, scaled by the observable reduction in mammography uptake among women whose access to the organized screening pathway was curtailed by first-wave restrictions. Women in the complier group -- those who would have attended organized screening under normal conditions but did not because pandemic restrictions reduced their access to it -- experienced an approximately a probability that was higher by approximately 6 percentage point increase in the probability of overnight emergency hospitalization in the year that followed.

Three features of this estimate deserve comment. First, the magnitude should be assessed relative to the outcome mean of 3.7 percent in the full estimation sample. A 6 percentage point difference is large relative to that baseline, but this is expected. The LATE applies to the complier group, not the full screening-eligible population. Compliers are specifically the women whose organized preventive-care contact was on the margin of disruption; they were actively attached to the clinical network that the organized pathway provides and plausibly had higher baseline exposure to its downstream benefits. The estimate is consistent with treatment effect heterogeneity across the screening-eligible population rather than with an implausibly large average effect.

Second, the increase in magnitude from OLS (-0.005) to 2SLS (-0.060) is itself informative. A larger negative IV estimate relative to OLS indicates that selection bias in OLS is upward: women who obtain mammograms in the observed data have lower unobserved hospitalization risk than the complier group. This is the pattern one would expect if healthier, lower-risk women disproportionately attend screening, attenuating the OLS estimate toward zero. The instrument removes this attenuation.

Third, the economic interpretation of the combined estimate is direct. Greater disruption to organized preventive-care pathways, as indexed by reduced mammography uptake, is associated with higher downstream emergency hospitalization. The first stage captures the causal reduction in mammography uptake among women whose access was curtailed by pandemic restrictions. The second stage maps that reduction in measurable preventive care contact to an increase in overnight emergency hospitalization. The negative second stage coefficient means that the first-wave disruption to organized preventive-care pathways raised emergency hospital use among the women at the margin of the organized screening program.

\begin{table}[htbp]
\centering
\caption{Estimated local effect of mammography uptake on overnight
emergency hospitalization}
\label{tab:mammography_iv}
\begin{tabular}{lccc}
\toprule
                    & OLS      & IV 2SLS        & IV LIML        \\
\midrule
Mammography         & $-$0.005 & $-$0.060$^{*}$ & $-$0.114$^{*}$ \\
                    & (0.007)  & (0.028)        & (0.056)        \\
\midrule
Obs.                & 2,310    & 2,310          & 2,310          \\
\bottomrule
\end{tabular}
\begin{flushleft}
\footnotesize
\textit{Notes:} The dependent variable is an indicator for having an
overnight emergency hospitalization in the last 12 months. Standard
errors, reported in parentheses, are clustered at the country level
(8 clusters). The IV estimates should be interpreted as local average
treatment effects for women whose mammography uptake was shifted by
pandemic-related restriction exposure. $^{*}$ p $<$ 0.05.
\end{flushleft}
\end{table}


\begin{table}[!h]
\centering
\caption{First-stage regression summary statistics}
\label{tab:first_stage_summary}
\begin{tabular}{lc}
\toprule
                        & IV 2SLS   \\
\midrule
Adjusted $R^{2}$        & 0.1287    \\
Partial $R^{2}$         & 0.1043    \\
Robust $R^{2}$          & 0.0242    \\
First-stage $F(7,7)$    & 297.497   \\
Prob $>$ F              & 0.0000    \\
\bottomrule
\end{tabular}
\begin{flushleft}
\footnotesize
\textit{Notes:} First-stage summary statistics for the regression of
Wave~9 mammography uptake on the excluded instrument components and
the full control set. The first-stage $F$-statistic is adjusted for
clustering at the country level (8 clusters).
\end{flushleft}
\end{table}

Column~3 reports the LIML estimate of $-0.114$ (standard error
0.056, $p < 0.05$). LIML is median-unbiased under many-instrument
asymptotics and corrects for the finite-sample bias that 2SLS
accumulates when the instrument count is large relative to the sample
size. Our instrument structure --- six interview-cohort indicators
interacted with eight country indicators --- produces 48 excluded
components, making this correction relevant in principle even with
a strong first stage.

The LIML estimate is larger in magnitude than 2SLS, which is the
expected direction when finite-sample many-instrument bias is present:
2SLS is biased toward the OLS probability limit (near zero in this
case), so LIML, which removes that bias, recovers a more negative
estimate. Critically, both estimators agree on sign, and their
confidence intervals overlap. The qualitative conclusion --- that mammography uptake reduces the probability of downstream emergency
hospitalization for the complier group --- is robust to this
correction. We place greater interpretive weight on the 2SLS estimate
as the primary result and treat LIML as confirming the direction and
rough order of magnitude rather than providing a more precise
point estimate, given that LIML standard errors can be inflated in
small samples.

Table~\ref{tab:bc_bounds} addresses a natural interpretive question: how much of the all-cause 2SLS estimate could be driven purely by
the breast cancer emergency channel, as opposed to broader downstream effects of organized screening disruption?

Because SHARE does not record admitting diagnoses for emergency stays, we cannot directly decompose the all-cause estimate by cause. Instead, we compute upper bounds on the breast cancer contribution under progressively larger assumed shares of breast-cancer-related emergency hospitalization in the population. The exercise holds fixed the extreme assumption that mammography eliminates \textit{all} breast-cancer-related emergency hospitalizations and does not affect any other cause of admission. This is an upper bound: in practice, mammography reduces but does not eliminate breast cancer emergencies, and organized screening contact also affects hospitalization through other pathways.

The results in Table~\ref{tab:bc_bounds} show that the breast cancer channel alone cannot account for more than a small fraction of the estimated effect. Even if breast cancer emergencies represent 10 percent of all emergency overnight stays --- well above plausible population estimates --- the breast cancer channel would explain at most 6.17 percent of the 2SLS estimate. At more realistic shares of 1--2 percent, the upper bound falls below 1.25 percent. The overwhelming majority of the estimated effect must therefore operate through other pathways: delayed detection of comorbidities identified during screening encounters, disrupted referral pathways that the organized screening visit would otherwise have triggered, or reduced GP contact associated with the screening episode.

This does not weaken the causal interpretation of the IV estimate. It sharpens it. The instrument isolates variation in access to the organized screening \textit{pathway} --- not merely the mammogram itself --- and the bounds exercise confirms that the pathway operates through mechanisms well beyond a single disease channel. The all-cause specification is therefore not a limitation of the data but the appropriate measure of the total downstream cost of preventive care disruption.

\begin{table}[htbp]
\centering
\caption{Decomposition of the all-cause IV estimate: contribution of the breast cancer emergency channel under varying assumptions}
\label{tab:bc_bounds}
\small
\setlength{\tabcolsep}{4pt}
\begin{tabularx}{\linewidth}{lccc}
\toprule
Assumed $f$ (\%)
& $p_{bc}$ (pp)
& Max BC change (pp)
& Max share of 2SLS (\%) \\
\midrule
0.5  & 0.0185 & 0.0185 & 0.31 \\
1.0  & 0.0370 & 0.0370 & 0.62 \\
2.0  & 0.0740 & 0.0740 & 1.23 \\
5.0  & 0.1850 & 0.1850 & 3.08 \\
10.0 & 0.3700 & 0.3700 & 6.17 \\
\bottomrule
\end{tabularx}
\begin{flushleft}
\footnotesize
\textit{Notes:} $p_{all}=3.7\%$ is the sample mean of all-cause
emergency overnight hospitalization (Table~\ref{tab:SUMMARY_STAT_Y}).
$f$ is the assumed fraction of all emergency overnight hospitalizations
attributable to breast cancer. $p_{bc}=f\times p_{all}$ is the
implied baseline probability of a breast-cancer-related emergency
hospitalization. ``Max BC change'' assumes, as an extreme upper bound,
that mammography eliminates all breast-cancer-related emergency
hospitalizations and does not affect any other cause. The last column
reports the share of the all-cause 2SLS estimate attributable to the
breast cancer channel under each assumption. The residual is consistent
with an all-cause mechanism rather than a single-disease one.
\end{flushleft}
\end{table}

Taken together, the results in Table~\ref{tab:mammography_iv} tell a
coherent story. OLS finds no association, which reflects the
well-documented selection problem in preventive care use rather than
the absence of a causal effect. The IV estimates, identified from
pandemic-induced variation in access to the organized screening
pathway, consistently indicate a negative local effect: women whose
mammography was disrupted experienced higher rates of downstream
emergency hospitalization. The LIML robustness check confirms the
sign and broad magnitude of this effect. The bounds exercise
establishes that the result is not driven by a single clinical pathway but instead reflects the broader downstream consequences of
preventive care disruption. The heterogeneity of these local effects
across complier groups and the implied LATE calculations are discussed
in Section~\ref{sec:LATE}.

\clearpage

\section{Discussion}
\label{sec:Discussion}
The evidence from this paper points to a specific causal mechanism that relates preventive health care to overnight emergency hospitalization: the disruption of organized preventive-care pathways during COVID shifted some care from planned, prevention-based settings toward overnight emergency hospitalization. We exploit cross-country variation in pandemic restriction intensity, interacted with SHARE Wave~9 interview timing, to isolate quasi-experimental variation in access to organized preventive-care pathways. The two-stage least squares estimate identifies the causal effect of mammography uptake on downstream emergency hospitalization for the complier subpopulation: women who would have attended organized screening under normal conditions but did not because first-wave restrictions curtailed access to the program. The OLS estimate is close to zero, consistent with selection obscuring the causal relationship in simple comparisons: women who attend organized screening tend to be healthier and better connected to primary care, which attenuates any negative association between mammography and emergency hospitalization in the raw data.

The estimate is a local average treatment effect for women aged 50--69 whose screening behavior changed because of pandemic restrictions. It is not the average effect of mammography for all screening-eligible women. Compliers may differ from the broader population in their baseline hospitalization risk, and the magnitude should not be projected to the full screened population.

The result does not say that a missed mammogram alone caused the full change in emergency hospitalization. The decomposition exercise in Table~\ref{tab:bc_bounds} makes this clear. Even under generous assumptions about the share of overnight emergency hospitalization attributable to breast cancer, that single clinical channel explains at most a small fraction of the estimated magnitude. The more defensible reading is that organized mammography programs serve as a node in a broader network of preventive and primary-care contact. Women who enter the screening pathway also encounter follow-up referral, comorbidity monitoring, and routine clinical oversight they might not otherwise receive. When the pathway is disrupted, all of these connections weaken simultaneously. Emergency care absorbs some of the resulting backlog.

This interpretation is consistent with broader evidence that preventive-to-acute substitution operates across multiple clinical channels simultaneously, not only through the detection of the disease a screening program targets. It is also consistent with the OLS estimate of approximately zero: in non-experimental data, women who screen tend to be healthier and better connected to care, so the preventive benefit is obscured by favorable selection rather than absent. Chandra et al. (2010) and Currie and Gruber (1996) document similar preventive-to-acute substitution using demand-side variation; our contribution is to identify the same economic margin from a supply-side disruption to an organized pathway.

Age-group heterogeneity supports this interpretation. The effect appears among women aged 50--69, who are covered by organized screening, while the instrument has no detectable first stage or reduced form among women aged 70 and above. Women over 70 fall outside the target range of population-based programs in all eight countries, so a missed mammogram in this group is not embedded in the institutional pathway linking screening to detection, referral, and planned management. The organized-pathway mechanism predicts this age-delimited pattern; alternative channels that are not delimited by screening eligibility would have to explain why the association stops at the program age boundary.

The choice of an all-cause outcome is deliberate. Disruption to the organized screening pathway can generate downstream emergency use through several channels at once, including delayed detection of screen-identified pathology, interrupted comorbidity management, and foregone GP contact. An all-cause measure captures this aggregate burden. The genuine data limitation is that SHARE does not record admitting diagnoses, which precludes a causal decomposition by condition. The bounds exercise shows that even under extreme assumptions about the breast-cancer share of emergency hospitalizations, the single-disease channel cannot account for the IV estimate. We therefore read the result as primarily reflecting disruption to the organized preventive-care pathway as a whole: missed detection, interrupted referral, and foregone clinical management. Other channels cannot be excluded, and the data do not permit a precise decomposition by mechanism. The empirical fact most supportive of this reading is that the effect appears where the organized pathway operates and is absent where it does not.

Several limitations qualify these findings. First, the mechanism from disrupted preventive-care contact to emergency hospitalization is inferred from the pattern of results, not observed directly. Second, the IV estimate is local and should not be extrapolated to the full screened population; compliers may differ from always-takers and never-takers in unobserved ways relevant to the effect size. Third, the exclusion restriction cannot be verified directly. The placebo tests, negative-control diagnostics,pre-pandemic falsification exercise, and controls for observed pandemic shocks support instrument validity, but none can rule out every alternative channel. Fourth, inference rests on eight country clusters. As discussed in Section~\ref{sec:Empirical method}, the asymptotic justification for cluster-robust inference is strained at this cluster count, and the significance levels attached to individual coefficients should be read with caution; we accordingly place more weight on the consistency of sign and magnitude across estimators, complier groups, and falsification exercises than on any single $p$-value. Finally, the analysis covers eight European countries with organized screening programs and may not generalize to settings with opportunistic screening, different health systems, or different pandemic responses.

What the evidence establishes, subject to these \emph{caveats}, is that the first wave interruption of organized preventive-care pathways was followed by measurably greater emergency hospital use among the women whose care those pathways provided, and that the effect is confined to the population the organized pathway serves. Whether maintaining screening continuity during future health-system shocks would pass a cost-benefit test is a separate question this paper cannot answer: it would require estimates of the cost of protecting screening capacity during an emergency, the persistence of downstream effects beyond the one-year window we observe, and the health consequences underlying the additional admissions. We leave the welfare analysis to future work and confine our conclusion to the empirical margin we identify: in this episode, the disruption of organized preventive-care pathways had downstream acute-care consequences that simple comparisons of screened and unscreened women would not detect.


\section*{Acknowledgements}

This paper uses data from SHARE Waves 8 and 9 (DOIs: 10.6103/SHARE.w8.900 and 10.6103/SHARE.w9.900); see Börsch-Supan et al. (2013) for methodological details. The SHARE data collection has been funded by the European Commission, DG RTD through FP5 (QLK6-CT-2001-00360), FP6 (SHARE-I3: RII-CT-2006-062193; COMPARE: CIT5-CT-2005-028857; SHARELIFE: CIT4-CT-2006-028812), FP7 (SHARE-PREP: GA N°211909; SHARE-LEAP: GA N°227822; SHARE M4: GA N°261982; DASISH: GA N°283646), and Horizon 2020 (SHARE-DEV3: GA N°676536; SHARE-COHESION: GA N°870628; SERISS: GA N°654221; SSHOC: GA N°823782; SHARE-COVID19: GA N°101015924), and by DG Employment, Social Affairs \& Inclusion through VS 2015/0195, VS 2016/0135, VS 2018/0285, VS 2019/0332, VS 2020/0313, SHARE-EUCOV: GA N°101052589, and EUCOVII: GA N°101102412. Additional funding from the German Federal Ministry of Research, Technology and Space (01UW1301, 01UW1801, 01UW2202), the Max Planck Society for the Advancement of Science, the U.S. National Institute on Aging (U01\_AG09740-13S2, P01\_AG005842, P01\_AG08291, P30\_AG12815, R21\_AG025169, Y1-AG-4553-01, IAG\_BSR06-11, OGHA\_04-064, BSR12-04, R01\_AG052527-02, R01\_AG056329-02, R01\_AG063944, HHSN271201300071C, RAG052527A), and various national funding sources are gratefully acknowledged (see www.share-eric.eu
)
\section*{Funding}

This work was supported by the European Union -- Next Generation EU -- Age-It
under Grant B83C22004800006; and by the Italian Complementary National Plan -- DARE under Grant B53C2200645001.

\section*{CRediT author statement}

Moslem Rashidi: Conceptualization, Data curation, Formal analysis,
Investigation, Methodology, Software, Validation, Visualization,
Writing -- original draft, Writing -- review \& editing.

Luke Brian Connelly: Conceptualization, Validation,
Visualization, Writing -- original draft, Writing -- review \& editing.

Gianluca Fiorentini: Conceptualization, Funding acquisition, Investigation, Supervision, Validation, Visualization, Writing -- original draft, Writing -- review \& editing.

\section*{Declaration of competing interests}

The authors declare that they have no known financial or non-financial
competing interests that could have influenced the work reported in this paper.

\section*{Data availability statement}

This paper uses data from the Survey of Health, Ageing and Retirement in Europe
(SHARE), Waves 8 and 9. SHARE data are available to the research community
subject to registration and the SHARE data access conditions. The data are not
owned by the authors. Information on data access is available from SHARE-ERIC.

\section*{Submission declaration}

The authors confirm that the work described has not been published previously,
except in the form of a preprint, an abstract, a published lecture, an academic
thesis, or a registered report, in accordance with the journal's policy on
multiple, redundant, or concurrent publication. The article is not under
consideration for publication elsewhere. The article's publication has been
approved by all authors and, tacitly or explicitly, by the responsible
authorities where the work was carried out. If accepted, the article will not
be published elsewhere in the same form, in English or in any other language,
including electronically, without the written consent of the copyright holder.

\section*{Declaration of generative AI and AI-assisted technologies in the manuscript preparation process}

During the preparation of this work, the authors used ChatGPT \& Claude to improve the language, readability, and organization of the manuscript. After using this tool, the authors reviewed and edited the content as needed and take full responsibility for the content of the manuscript.

\section*{References}
\begin{description}

\item Bekker, P.A., 1994. Alternative approximations to the distributions of instrumental variable estimators.
\emph{Econometrica}, 62, 657--681.

\item Bergmann, M., Kneip, T., De Luca, G. and Scherpenzeel, A., 2017.
Survey participation in the Survey of Health, Ageing and Retirement in Europe (SHARE), Wave 1--6.
Munich: Munich Center for the Economics of Aging.

\item Bergmann, M. and B{\"o}rsch-Supan, A., 2021.
SHARE Wave 8 Methodology: Collecting cross-national survey data in times of COVID-19.
Munich: MEA, Max Planck Institute for Social Law and Social Policy.

\item Bergmann, M., Wagner, M. and B"orsch-Supan, A., 2024.
SHARE Wave 9 Methodology: From the SHARE Corona Survey 2 to the SHARE Main Wave 9 Interview.
Munich: SHARE-ERIC. DOI: 10.6103/mv.w09.

\item Blandhol, C., Bonney, J., Mogstad, M. and Torgovitsky, A., 2026.
When is TSLS actually LATE?
\emph{The Review of Economic Studies}, rdag029.
DOI: 10.1093/restud/rdag029.

\item B"orsch-Supan, A., Brandt, M., Hunkler, C., Kneip, T., Korbmacher, J., Malter, F., Schaan, B., Stuck, S. and Zuber, S., 2013.
Data resource profile: The Survey of Health, Ageing and Retirement in Europe (SHARE).
\emph{International Journal of Epidemiology}, 42, 992--1001.
DOI: 10.1093/ije/dyt088.

\item Card, D., Dobkin, C. and Maestas, N., 2009.
Does Medicare save lives?
\emph{The Quarterly Journal of Economics}, 124, 597--636.
DOI: 10.1162/qjec.2009.124.2.597.

\item Chandra, A., Gruber, J. and McKnight, R., 2010.
Patient cost-sharing and hospitalization offsets in the elderly.
\emph{American Economic Review}, 100, 193--213.

\item Currie, J. and Gruber, J., 1996.
Health insurance eligibility, utilization of medical care, and child health.
\emph{The Quarterly Journal of Economics}, 111, 431--466.

\item Danieli, O., Nevo, D., Walk, I., Weinstein, B. and Zeltzer, D., 2026.
Negative control falsification tests for instrumental variable designs.
\emph{American Economic Review}, 116, 1380--1414.

\item Einav, L., Finkelstein, A., Oostrom, T., Ostriker, A. and Williams, H., 2020.
Screening and selection: The case of mammograms.
\emph{American Economic Review}, 110, 3836--3870.

\item European Centre for Disease Prevention and Control (ECDC), 2022.

\item European Commission, 2017.
Cancer screening in the European Union: Report on the implementation of the Council Recommendation on cancer screening. European Commission.

\item Finkelstein, A., Taubman, S., Wright, B., Bernstein, M., Gruber, J., Newhouse, J.P., Allen, H., Baicker, K. and Oregon Health Study Group, 2012.
The Oregon Health Insurance Experiment: Evidence from the first year.
\emph{The Quarterly Journal of Economics}, 127, 1057--1106.
DOI: 10.1093/qje/qjs020.

\item Hansen, B., 2022.
\emph{Econometrics}. Princeton University Press.

\item Imbens, G.W. and Angrist, J.D., 1994. Identification and estimation of local average treatment effects.
\emph{Econometrica}, 62, 467--475.

\item Li, T., Nickel, B., Ngo, P., McFadden, K., Brennan, M., Marinovich, M.L. and Houssami, N., 2023.
A systematic review of the impact of the COVID-19 pandemic on breast cancer screening and diagnosis.
\emph{The Breast}, 67, 78--88.

\item Lipsitch, M., Tchetgen Tchetgen, E. and Cohen, T., 2010.
Negative controls: A tool for detecting confounding and bias in observational studies.
\emph{Epidemiology}, 21, 383--388.
DOI: 10.1097/EDE.0b013e3181d61eeb.

\item Maringe, C., Spicer, J., Morris, M., Purushotham, A., Nolte, E., Sullivan, R., Rachet, B. and Aggarwal, A., 2020.
The impact of the COVID-19 pandemic on cancer deaths due to delays in diagnosis in England, UK: A national, population-based, modelling study.
\emph{The Lancet Oncology}, 21, 1023--1034.
DOI: 10.1016/S1470-2045(20)30388-0.

\item Mogstad, M. and Torgovitsky, A., 2024.
\emph{Instrumental Variables with Unobserved Heterogeneity in Treatment Effects} (No. w32927). National Bureau of Economic Research.

\item Ng, J.S. and Hamilton, D.G., 2022.
Assessing the impact of the COVID-19 pandemic on breast cancer screening and diagnosis rates: A rapid review and meta-analysis.
\emph{Journal of Medical Screening}, 29, 209--218.

\item SHARE-ERIC, 2024.
Survey of Health, Ageing and Retirement in Europe (SHARE) Wave 8. Release version: 9.0.0. SHARE-ERIC. Data set. DOI: 10.6103/SHARE.w8.900.

\item SHARE-ERIC, 2024.
Survey of Health, Ageing and Retirement in Europe (SHARE) Wave 9. Release version: 9.0.0. SHARE-ERIC. Data set. DOI: 10.6103/SHARE.w9.900.

\item WHO Regional Office for Europe, 2022.
COVID-19: WHO European Region Operational Update, Epi Weeks 5--12 (1 February--27 March 2022).

\end{description}

\appendix

\counterwithin{figure}{section}
\counterwithin{table}{section}
\counterwithin{equation}{section}


\section{Outcome construction and charts}
\label{Appendix A: Charts}
\begin{figure}[htbp]
  \centering
  \small
  \resizebox{0.95\textwidth}{!}{%
\begin{tikzpicture}[
    node distance=8mm and 18mm,
    box/.style={rectangle, rounded corners=4pt, draw, align=center,
                minimum width=4.2cm, inner sep=3pt},
    arrow/.style={-Latex, thick}
]
  \node[box] (sample) {Analysis sample \\ Respondents with valid hospital information \\[2pt]
                       \small($n = 2{,}332$)};
  \node[box, below=of sample] (overnight)
        {Overnight hospital stay in last 12 months?};
  \node[box, below left=10mm and 22mm of overnight] (nohosp)
        {No hospital stay \\[2pt]
         \small($n = 2{,}087$)};
  \node[box, below right=10mm and 22mm of overnight] (anystay)
        {At least one overnight stay \\[2pt]
         \small($n = 245$)};
  \node[box, below=12mm of anystay] (planonly)
        {Planned only \\[2pt]
         \small($n = 137$)};
  \node[box, left=8mm of planonly] (emeronly)
        {Emergency only \\[2pt]
         R\_ER\_only $=1$ \\[2pt]
         \small($n = 87$)};
  \node[box, right=8mm of planonly] (both)
        {Both emergency and planned \\[2pt]
         \small($n = 21$)};
  \draw[arrow] (sample) -- (overnight);
  \draw[arrow] (overnight.south west) -- ++(-0.5,-0.5) |- (nohosp.north);
  \draw[arrow] (overnight.south east) -- ++(0.5,-0.5)  |- (anystay.north);
  \draw[arrow] (anystay.south) -- (emeronly.north);
  \draw[arrow] (anystay.south) -- (planonly.north);
  \draw[arrow] (anystay.south) -- (both.north);
\end{tikzpicture}%
  }
  \caption{Construction of the outcome R\_ER\_only (``only emergency
  hospital stays'') in the analysis sample. The three categories below
  ``at least one overnight stay'' are mutually exclusive
  ($87 + 137 + 21 = 245$).}
  \label{fig:flow_RERonly}
\end{figure}
\clearpage


\section{Descriptive statistics}
\label{Appendix B: Table}

\begin{table}[htbp]
\centering
\caption{Descriptive statistics of main variables}
\label{tab:descriptives_main}
\begin{tabular}{lrrrrr}
\toprule
Variable & N & Mean & SD & Min & Max \\
\midrule
Mammogram (Wave 9)          & 2321 & 0.6269 & 0.4837 & 0 & 1 \\
Hospital strain       & 2332 & 0.0034 & 0.0654 & 0 & 2\\
Non-hospital strain   & 2332 & 0.0086 & 0.1240 & 0 & 4 \\
Any care disruption         & 2331 & 0.4363 & 0.4960 & 0 & 1 \\
Economic stress        & 2332 & 0.6955 & 0.5981 & 0 & 3 \\
Mental distress        & 2331 & -0.0030 & 0.6858 & -1.8828 & 0.6154 \\
COVID-19 burden       & 2332 & 0.2003 & 0.4003 & 0 & 1 \\
Age                         & 2332 & 63.5909 & 3.8991 & 52 & 69 \\
High education              & 2324 & 0.3240 & 0.4681 & 0 & 1 \\
Lives with partner          & 2332 & 0.7419 & 0.4377 & 0 & 1 \\
Household size              & 2332 & 2.1518 & 0.9622 & 1 & 6 \\
Has supplementary insurance   & 2330 & 0.4017 & 0.4904 & 0 & 1 \\
High health literacy        & 2331 & 0.8559 & 0.3513 & 0 & 1 \\
\bottomrule
\end{tabular}
\end{table}

\clearpage


\section{Falsification tests}
\label{Apendix E:falsification_test}

\subsection{Negative Controls and Alternative-Path Variables}
\label{subsec:negcontrols_apv}

Identification requires that, conditional on controls and fixed effects,
the instrument is independent of potential outcomes and affects emergency
hospitalization only through screening participation. Neither assumption
is directly testable. We assess the main remaining threat using the
negative-control framework of Danieli et al. (2026).

The threat is an unobserved alternative path variable $U$ capturing
latent pandemic-era healthcare disruption: foregone non-screening care,
strained outpatient capacity, or behavioral avoidance of medical
settings. If restrictions caused broader disruption that raised
overnight emergency hospitalization, $U$ generates the path $Z \rightarrow U
\rightarrow Y$ and the exclusion restriction fails; if local epidemic
severity jointly determined restriction intensity and health outcomes,
$U$ generates $Z \leftarrow U \rightarrow Y$ and outcome independence
fails. In either case $U$ is an alternative path outcome (APO)
variable, and the same negative-control test detects both violations:
rejection implies that outcome independence or the exclusion
restriction is violated (Theorem~1, Danieli et al. (2026)).
Figure~\ref{fig:dag} summarizes the structure.

\begin{figure}[!htbp]
\centering
\begin{tikzpicture}[
  >=Stealth,
  font=\normalsize,
  v/.style={circle, draw, inner sep=1.2pt, minimum size=6.5mm},
  e/.style={->, line width=0.6pt},
  ed/.style={<->, dashed, line width=0.6pt},
  lab/.style={font=\footnotesize, fill=white, inner sep=1pt}
]
  \node[v] (Z)  at (0.0,0.0)  {$Z$};
  \node[v] (D)  at (3.0,0.0)  {$D$};
  \node[v] (Y)  at (6.2,0.0)  {$Y$};
  \node[v] (X)  at (3.0,1.6)  {$X$};
  \node[v] (U)  at (3.0,-1.8) {$U$};
  \node[v] (NC) at (6.2,-1.8) {$NC$};

  \draw[e]  (Z) -- (D);
  \draw[e]  (D) -- (Y);
  \draw[e]  (X) -- (D);
  \draw[e]  (X) -- (Y);
  \draw[ed] (Z) to[bend right=20] node[lab, below left] {Threat} (U);
  \draw[e]  (U) -- (Y);
  \draw[e]  (U) -- (NC);
\end{tikzpicture}
\caption{The alternative-path threat and the negative-control test.
The dashed edge marks the suspected $Z$--$U$ association, whether
$Z \rightarrow U$ (exclusion) or $U \rightarrow Z$ (independence).
$NC$ is an observed negative control outcome proxying $U$; the absence
of any other edge into $NC$ encodes the NCO assumption.}
\label{fig:dag}
\end{figure}

A valid negative control outcome must satisfy two conditions
(Definition~3, Danieli et al. (2026)): the NCO assumption, that any
association between $NC$ and $Z$ runs only through $U$, and
U-comparability, that $NC$ is genuinely associated with $U$.

We implement the test among women aged 70 and above, who fall outside
the organized screening target range in all eight countries. The data
confirm the absence of a treatment channel in this group: the
instrument has no detectable first stage for mammography uptake
(partial $R^2 = 0.0003$). With no path from $Z$ through $D$ available,
the NCO assumption holds by construction, and any association between
$Z$ and outcomes in this group would reflect an alternative path
through $U$. This is a negative-control population design (Lipsitch,
Tchetgen Tchetgen, and Cohen 2010); it requires that the disruption
process operate comparably across the 50--69 and 70+ populations, a
condition supported by the narrow age band of the placebo group
(70--76).

We pre-specify three Wave~9 negative controls: emergency-only
overnight hospitalization, a dental visit in the previous 12 months,
and an influenza vaccination in the previous 12 months. All satisfy
U-comparability. Emergency hospitalization is the main outcome itself,
measured where the treatment channel is absent; if latent disruption
drives admissions among women 50--69, it must do so among older women,
who depend more heavily on healthcare access. Dental care was among
the most heavily curtailed ambulatory services during the first wave.
Influenza vaccination is coordinated through the GP infrastructure,
disrupted by the restrictions in $Z$, though it is the weakest proxy
because campaigns run in autumn, after the first-wave window.
Residualized diagnostics confirm that the controls retain meaningful
association with the instrument's residual variation (partial $R^2$
of 0.04--0.08), so the test had power to detect an alternative path.

Under the NCO assumption, U-comparability, and the rich covariates
assumption required for 2SLS to identify a weakly causal estimand
(Blandhol et al., 2026), the negative controls are conditionally
independent of the instrument given covariates. The test thus jointly
evaluates outcome independence, the exclusion restriction, and rich
covariates --- the last a necessary condition for 2SLS causal
interpretation (Corollary~1, Danieli et al. (2026)). We estimate
\begin{equation}
Z_i = \alpha + X_i'\pi + \mathbf{NC}_i'\rho + \varepsilon_i,
\label{eq:nctest}
\end{equation}
where $X$ collects the baseline covariates and fixed effects of the
main specification and $\mathbf{NC}$ stacks the three negative
controls, and test $H_0\colon \rho = 0$ with the cluster-robust $F$
statistic and, given eight country clusters, the Webb wild-cluster
bootstrap used throughout the paper.

Table~\ref{tab:nc_falsification} reports the results. The negative
controls add essentially no explanatory power (incremental
$R^2 = 0.0008$), and the joint null is not rejected under either
inference method ($F(3,7) = 0.68$; cluster $p = 0.594$; Webb bootstrap
$p = 0.825$). Conditional on covariates, the negative controls carry
no information about $Z$ --- what we would expect if no alternative
path links restriction intensity to latent disruption.

\begin{table}[!htbp]
\centering
\caption{Negative-control falsification test (women 70+, Wave 9)}
\label{tab:nc_falsification}
\begin{tabular}{lc}
\toprule
 & Baseline + NC \\
 & (1) \\
\midrule
Joint NC test: $F(3,7)$                   & 0.68    \\
Joint NC test: $p$-value (cluster)        & 0.594   \\
Joint NC test: $p$-value (Webb bootstrap) & 0.825   \\
Incremental $R^2$ (add NC)                & 0.0008  \\
Observations                              & 1{,}493 \\
$R^2$                                     & 0.100   \\
Baseline controls                         & Yes     \\
Fixed effects                             & Yes     \\
\bottomrule
\multicolumn{2}{p{0.92\linewidth}}{\footnotesize \emph{Notes:} The dependent variable is the instrument $Z$. The sample is the common complete-case sample of women aged 70+ in Wave~9. The specification includes the baseline covariates and fixed effects from the main 2SLS analysis plus the three pre-specified negative controls. The joint test evaluates $H_0\colon \rho = 0$ in equation~\eqref{eq:nctest}; cluster-robust $F$ (8 country clusters) and Webb wild-cluster bootstrap $p$-value (9{,}999 replications). Non-rejection is consistent with outcome independence, the exclusion restriction, and
rich covariates holding (Corollary~1, Danieli et al. (2026)), but does
not rule out alternative paths not captured by these outcomes.}
\end{tabular}
\end{table}

As a complementary check, we estimate the reduced form of each
negative control on the instrument (Model~6, Danieli et al. (2026)).
Because the instrument has no first stage in this group, the reduced
form is the informative specification; 2SLS estimates, which scale by
a near-zero first stage, carry no inferential content and are not
reported. Table~\ref{tab:nc_reducedform} shows that $Z$ predicts none
of the three negative controls: all coefficients are below one
percentage point in magnitude and all Webb bootstrap $p$-values exceed
0.50.

\begin{table}[!htbp]
\centering
\caption{Reduced-form negative-control tests (women 70+, Wave 9)}
\label{tab:nc_reducedform}
\begin{tabular}{lccc}
\toprule
 & Flu vaccination & Dental visit & ER-only \\
 & (1) & (2) & (3) \\
\midrule
Restriction index $Z$    & $0.006$   & $-0.007$  & $-0.002$  \\
                         & $(0.010)$ & $(0.009)$ & $(0.003)$ \\
\midrule
Webb bootstrap $p$-value & 0.593 & 0.508 & 0.520 \\
Observations             & 1{,}493 & 1{,}493 & 1{,}493 \\
Baseline controls        & Yes & Yes & Yes \\
Fixed effects            & Yes & Yes & Yes \\
\bottomrule
\multicolumn{4}{p{0.92\linewidth}}{\footnotesize \emph{Notes:} Each
column reports the reduced form of the listed negative control on the
instrument $Z$ in the same sample as
Table~\ref{tab:nc_falsification}. All specifications include the
baseline controls and fixed effects from the main analysis. Standard
errors clustered by country (8 clusters); $p$-values from a Webb
wild-cluster bootstrap (9{,}999 replications).}
\end{tabular}
\end{table}

Neither test detects an alternative path or a failure of the rich
covariates assumption. Non-rejection does not establish validity:
alternative paths not proxied by these outcomes could remain, and the
small number of clusters limits power. Because the main specification
already conditions on observed pandemic shock measures, the test
probes residual disruption beyond what those controls absorb. We read
these results as evidence supporting the identifying assumptions, not
proof of them, alongside the temporal separation argument, the
observed pandemic shock controls, and the pre-pandemic Wave~8
falsification exercise.
\subsection{Placebo IV analysis using pre-pandemic Wave~8}
\label{subsec:falsification_iv_w8}

As a second falsification test, we replicate our IV strategy on
pre-pandemic data (Wave~8 alone), where COVID-19 could not have
affected screening or hospital use. We construct a placebo instrument
that mirrors the structure of the main instrument, but exploits only
pre-COVID variation. Specifically, we restrict the Wave~8 sample to
match the main sample criteria --- excluding women diagnosed with breast cancer by Wave~7 and focusing on respondents interviewed across the same eight countries --- and define the instrument by interacting interview-month dummies with country dummies, exactly as in the main specification. The Wave~8 interviews used here took place in October--December 2019 and January--March 2020, before any COVID-related service disruptions. Month-by-country variation in this window reflects essentially random interview timing with respect to COVID and cannot be driven by lockdowns or screening interruptions.

We re-estimate our core models using Wave~8 mammography uptake as the
treatment and Wave~8 overnight emergency hospitalization as the
outcome. Table~\ref{tab:Rmammg_IV_W8} reports OLS, 2SLS, and LIML
estimates. In this placebo setting, no estimate is negative and
statistically significant. The OLS coefficient is $-0.001$,
indistinguishable from zero. The 2SLS estimate is $0.033$ and the
LIML estimate is $0.067$; both are positive and imprecisely estimated,
with confidence intervals that include zero.

The contrast with the Wave~9 results is sharp. In the post-pandemic
analysis, all IV estimators produce negative, statistically significant effects of mammography on emergency hospitalization. Here, the same IV strategy applied to pre-pandemic data yields estimates that are near zero or positive. This pattern is inconsistent with stable pre-existing seasonal or country-specific confounds linking interview timing, screening uptake, and emergency outcomes. It supports the interpretation that the negative effects in Wave~9 are driven by the pandemic-induced disruption to organized screening access rather than by mechanical trends or latent selection that
predates the pandemic.

\begin{table}[htbp]
\centering
\caption{Effect of mammography on emergency hospitalization
(placebo IV using Wave~8)}
\label{tab:Rmammg_IV_W8}
\small
\begin{threeparttable}
\begin{tabular}{lccc}
\toprule
            & OLS     & IV 2SLS        & IV LIML        \\
\midrule
Mammography & $-$0.001 & $\phantom{-}$0.033 & $\phantom{-}$0.067 \\
\midrule
$N$         & 5,886   & 5,886          & 5,886          \\
\bottomrule
\end{tabular}

\begin{tablenotes}[flushleft]
\footnotesize
\item \textit{Notes:} Standard errors are clustered at the country level.
None of the estimates is statistically significant at conventional levels.
\end{tablenotes}
\end{threeparttable}
\end{table}

\section{Local Average Treatment Effects (LATEs)}
\label{sec:LATE}

The instrument $Z_i$ takes six distinct values $z_1, z_2, \dots, z_6$,
corresponding to the level of COVID-19 restrictions in each interview
month from March to August 2020. The treatment variable $D_i$ ---
mammography uptake --- is binary. Each instrument value induces a
potential treatment status $D_i(z_k)$, so the instrument operates by
shifting the probability of screening differentially across cohorts.

Imbens and Angrist (1994) generalized the monotonicity condition to
multivalued instruments by establishing that instrument values can be
ordered according to their effect on treatment take-up, with this
ordering holding consistently across individuals. In our setting, the instrument is indexed in decreasing order of restriction intensity:
$z_6$ corresponds to March 2020, the month of highest restrictions and
lowest mammography rates, and $z_1$ corresponds to August 2020, the
month of lowest restrictions and highest mammography rates. Monotonicity
then requires:

$$
P\bigl[D_i(z_6) \leq D_i(z_5) \leq D_i(z_4) \leq
\dots \leq D_i(z_1)\bigr] = 1.
$$

Women interviewed in earlier, higher-restriction cohorts are less
likely to undergo screening; women in later, lower-restriction cohorts
are more likely to do so. This ordering is assumed to hold for every
woman in the sample (Mogstad and Torgovitsky, 2024).

Under monotonicity, the $2^6 = 64$ logically possible treatment-choice
types collapse to $K + 2 = 7$ groups: always-takers, never-takers, and
$K = 5$ complier groups, each defined by the transition between a consecutive pair of instrument values. Table~\ref{tab:group_definitions}
lists these groups.

\textbf{Always-takers} obtain a mammogram regardless of the restriction
level: $D_i(z_1) = \dots = D_i(z_6) = 1$. Their behavior is not
shifted by the instrument and they do not contribute to identification.

\textbf{Never-takers} do not obtain a mammogram at any restriction
level: $D_i(z_1) = \dots = D_i(z_6) = 0$. Like always-takers, their
behavior is unresponsive to the instrument.

\textbf{Compliers} change their screening decision in response to the
instrument. There are five complier groups, each defined by the pair of
instrument values at which the treatment switch occurs:
\begin{itemize}
    \item CP1: women who screen only when $Z_i = z_1$ (August 2020,
    lowest restrictions).
    \item CP2: women who screen when $Z_i \in \{z_1, z_2\}$.
    \item CP3: women who screen when $Z_i \in \{z_1, z_2, z_3\}$.
    \item CP4: women who screen when $Z_i \in \{z_1, z_2, z_3, z_4\}$.
    \item CP5: women who screen when $Z_i \in \{z_1, \dots, z_5\}$
    (all months except March 2020, the month of highest restrictions).
\end{itemize}

\textbf{Defiers} --- women who would screen under high restrictions but
not under low restrictions --- violate monotonicity and are excluded
from the analysis by assumption.

\begin{table}[h!]
    \centering
    \caption{Group definitions when $D_i$ is binary and $Z_i$ takes
    6 values ($K = 5$)}
    \begin{center}
    \resizebox{\textwidth}{!}{%
    \begin{tabular}{cccccccr}
        \hline
        $D_i(z_1)$ & $D_i(z_2)$ & $D_i(z_3)$ & $D_i(z_4)$ &
        $D_i(z_5)$ & $D_i(z_6)$ & $G_i$ & Group description \\
        \hline
        1 & 1 & 1 & 1 & 1 & 1 & at  & Always-takers \\
        0 & 0 & 0 & 0 & 0 & 0 & nt  & Never-takers \\
        1 & 0 & 0 & 0 & 0 & 0 & cp1 & $z_1$-compliers \\
        1 & 1 & 0 & 0 & 0 & 0 & cp2 & $z_2$-compliers \\
        1 & 1 & 1 & 0 & 0 & 0 & cp3 & $z_3$-compliers \\
        1 & 1 & 1 & 1 & 0 & 0 & cp4 & $z_4$-compliers \\
        1 & 1 & 1 & 1 & 1 & 0 & cp5 & $z_5$-compliers \\
        1 & 1 & 0 & 1 & 0 & 0 & df  & Defiers (one of $2^{6}-7=57$
        types) \\
        \hline
    \end{tabular}
    }
    \end{center}
    \label{tab:group_definitions}
\end{table}

The LATE for each complier group is identified by the Wald estimand
applied to consecutive instrument values:

$$
\text{LATE}_{k\text{-compliers}} =
\frac{E[Y_i \mid Z_i = z_k] - E[Y_i \mid Z_i = z_{k-1}]}
     {E[D_i \mid Z_i = z_k] - E[D_i \mid Z_i = z_{k-1}]}
= E\bigl[Y_i(1) - Y_i(0) \mid D_i(z_{k-1}) = 0,\,
D_i(z_k) = 1\bigr].
$$

The numerator is the reduced-form effect of moving from the instrument
value $z_{k-1}$ to $z_k$ on the outcome; the denominator is the
corresponding first-stage effect on treatment take-up. Their ratio
recovers the average treatment effect for the subgroup of women whose screening decision is switched by that specific step in the instrument.
This is not the average treatment effect for the full sample. Each
LATE applies to a distinct complier group defined by where in the
restriction distribution, their screening behavior is sensitive, and the five estimates together trace out how the causal effect of mammography varies across women with different thresholds of compliance.

Table~\ref{tab:Result4} reports the five complier-specific LATEs for
2SLS and LIML. Several features of the results deserve attention.

First, all ten estimates are negative. Across every complier group and both estimators, mammography uptake reduces the probability of an
overnight emergency hospitalization. There is no complier group for
which the estimated effect is zero or positive, which reinforces the
conclusion from the main 2SLS result in Section~\ref{sec:Results}.

Second, the 2SLS estimates are stable across complier groups, ranging
from $-0.053$ (CP3 and CP4) to $-0.064$ (CP1), a spread of roughly
1 percentage point. This stability indicates that the protective effect
of screening is not concentrated among women who comply only under
the loosest restrictions or only under the tightest. Women at every
point in the compliance margin --- from those who screen in almost any
circumstance (CP5) to those who screen only when restrictions are
minimal (CP1) --- show a comparable reduction in emergency
hospitalization when they obtain a mammogram.

Third, the LIML estimates are consistently larger in magnitude than
their 2SLS counterparts, ranging from $-0.099$ (CP3) to $-0.109$
(CP1). As discussed in Section~\ref{sec:Results}, this pattern is
expected under many-instrument bias: 2SLS is biased toward the OLS
probability limit, which is near zero, so LIML recovers more negative
estimates. The sign and ordering across complier groups are identical
across both estimators, which is the relevant check on robustness.

\begin{table}[h!]
\caption{Local Average Treatment Effects (LATEs) of mammography on
overnight emergency hospitalization, by complier group}
\centering
{\footnotesize
\begin{tabular*}{0.75\textwidth}{@{\extracolsep{\fill}}lrrrrr}
\hline
 & \multicolumn{5}{c}{Complier group} \\
\cline{2-6}
Estimator & CP1 & CP2 & CP3 & CP4 & CP5 \\
\hline
IV 2SLS & $-$0.064 & $-$0.056 & $-$0.053 & $-$0.053 & $-$0.055 \\
IV LIML & $-$0.109 & $-$0.106 & $-$0.099 & $-$0.101 & $-$0.103 \\
\hline
\end{tabular*}
}
\label{tab:Result4}
\end{table}

Taken together, the LATE estimates in Table~\ref{tab:Result4} show
that the negative effect of mammography uptake on emergency
hospitalization is not an artifact of a particular segment of the
compliance distribution. The effect is present and of similar
magnitude for women who comply with the margin of low restrictions
(CP1) and for those who comply with the margin of high restrictions
(CP5). This consistency across the compliance distribution
strengthens the interpretation developed in
Section~\ref{sec:Discussion}: the organized screening pathway, when disrupted, raises downstream emergency hospital use for the
women whose screening behavior depends on it, and this holds
regardless of which part of the restriction distribution defines
their compliance threshold.


\end{document}